# Optimum Resource Allocation in 6G Optical Wireless Communication Systems

Osama Zwaid Alsulami
*School of Electronic and Electrical Engineering*
*University of Leeds*
Leeds, United Kingdom
ml15ozma@leeds.ac.uk

Amal A. Alahmadi
*School of Electronic and Electrical Engineering*
*University of Leeds*
Leeds, United Kingdom
elaaal@leeds.ac.uk

Sarah O. M. Saeed
*School of Electronic and Electrical Engineering*
*University of Leeds*
Leeds, United Kingdom
elsoms@leeds.ac.uk

Sanaa Hamid Mohamed
*School of Electronic and Electrical Engineering*
*University of Leeds*
Leeds, United Kingdom
elshm@leeds.ac.uk

T. E. H. El-Gorashi
*School of Electronic and Electrical Engineering*
*University of Leeds*
Leeds, United Kingdom
t.e.h.elgorashi@leeds.ac.uk

Mohammed T. Alresheedi
*Department of Electrical Engineering*
*King Saud University*
Riyadh, Saudi Arabia
malresheedi@ksu.edu.sa

Jaafar M. H. Elmirghani
*School of Electronic and Electrical Engineering*
*University of Leeds*
Leeds, United Kingdom
j.m.h.elmirghani@leeds.ac.uk

*Abstract*— Optical wireless communication (OWC) systems are a promising communication technology that can provide high data rates into the tens of Tb/s and can support multiple users at the same time. This paper investigates the optimum allocation of resources in wavelength division multiple access (WDMA) OWC systems to support multiple users. A mixed-integer linear programming (MILP) model is developed to optimise the resource allocation. Two types of receivers are examined, an angle diversity receiver (ADR) and an imaging receiver (ImR). The ImR can support high data rates up to 14 Gbps for each user with a higher signal to interference and noise ratio (SINR). The ImR receiver provides a better result compared to the ADR in term of channel bandwidth, SINR and data rate. Given the highly directional nature of light, the space dimension can be exploited to enable the co-existence of multiple, spatially separated, links and thus aggregate data rates into the Tb/s. We have considered a visible light communication (VLC) setting with four wavelengths per access point (red, green, yellow and blue). In the infrared spectrum, commercial sources exist that can support up to 100 wavelengths, significantly increasing the system aggregate capacity. Other orthogonal domains can be exploited to lead to higher capacities in these future systems in 6G and beyond.

## I. INTRODUCTION

Since the number of wireless communication users has increased dramatically, the demand for larger bandwidths and higher data rates has grown. Broadband radio frequency (RF) technology— the current wireless technology widely utilised in indoor environments—has a number of limitations. The scarce and congested radio spectrum is one of these limitations, which may cause limited channel capacity and low transmission rates. As a result, a number of approaches, such as smart antennas, advanced modulation, and multiple inputs and multiple outputs (MIMO) systems have been proposed to improve the use of the radio spectrum and to overcome these limitations [1], [2]. However, achieving data rates above 10 Gbps, for each user, is challenging when using the congested radio spectrum. By 2021, Cisco expects the Internet traffic to increase 27 times compared to its 2017 levels [3]. Thus, the increasing demand for high data rates is driving researchers to seek alternative parts of the spectrum, other than the 300 GHz of radio spectrum currently in use and/or proposed for near future use. The optical spectrum is a potential solution. It offers excellent channel characteristics in the indoor environment (where over 80% of traffic sarts and terminates indoor), abundant bandwidth and established low cost components [4]–[11]. Recently, many studies have shown that video, data, and voice can be transmitted through OWC systems at high data rates of up to 25 Gbps per user, and beyond in indoor environments [10]–[21]. Different transmitter and receiver configurations have been investigated to help decrease the delay spread and increase the signal-to-noise ratio [16], [22]–[29]. Multiple access techniques borrowed from RF systems, including methods that use the time, wavelength or code domains can help support multiple users in OWC systems. However, when the number of users increases, efficient utilisation of resources is required to avoid degradation in the signal quality.

This paper proposes an indoor multiple access OWC system that uses wavelength division multiple access (WDMA) to support multiple users, and unlike previous studies [7], [30]–[34], this work optimises the allocation of wavelengths and access points to users to maximise the system aggregate capacity, with SINR used as the metric of interest. The rest of this paper is organised as follows: Section II describes the system configuration including the room configuration, receiver types and MILP optimisation model. Section III illustrates the simulation setup and results and finally Section IV presents the conclusions.

## II. SYSTEM CONFIGURATION

An empty room that has no windows or doors was utilised in the simulation. We studied the impact of doors, windows and furniture in previous work [16], [35], with the empty room setting being used as a reference scenario. The optical indoor channel was modelled using ray tracing following [36] and [35]. Up to second order reflections were considered in this work as the third and higher order reflections have a small effect on the received power [36]. Surfaces inside the room are divided into small equals areas with size $dA$, and a reflection coefficient of $\rho$. It has been shown in [37] that plaster walls reflect light rays in the form of a Lambertian pattern. Thus, surfaces inside the room (ceiling, walls and floor) were modelled as Lambertian reflectors. Elements in each surface act as a secondary emitter that reflect the light rays in the form of a Lambertian pattern with $n$ (emission order of the Lambertain pattren) equal to 1. The area of these elements plays a significant role in the resolution of the results. Higher resolution results are produced, when the area of the elements is reduced. However, reducing the elements area comes at the cost of increased computation time when determining the channel properties. Table 1 shows the simulation parameters. All the communication links operate above the communication floor (CF) which is set at 1 m above the floor as shown in Fig. 1.

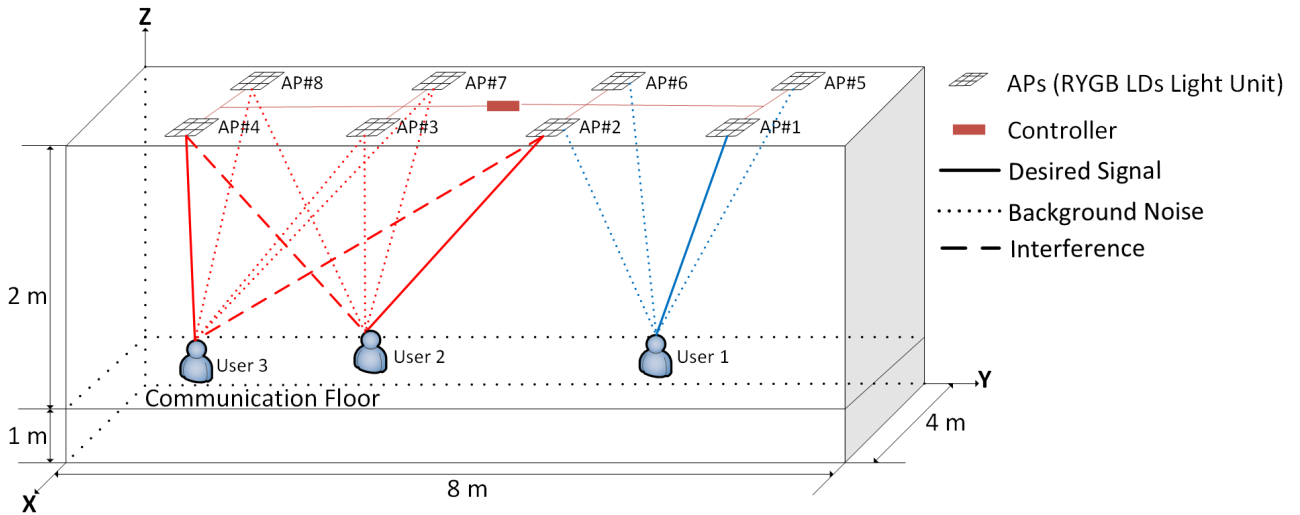

**Fig. 1.** System Configuration

Eight access points (AP) have been used in this work for illumination and communication. In addition, two types of optical receiver have been evaluated in this work. An angle diversity receiver (ADR) similar to [10] was compared with an imaging receiver (ImR) similar to [12]. Table 1 states the parameters of the AP and both types of receivers.

### III. SIMULATION SETUP AND RESULTS

In this work, two different 8-users scenarios were considered (see Tables 2 and 3 for more information). The first scenario was chosen as the worst scenario where each group of four users are clustered under an AP (each AP has four wavelengths in visible light communication: Red, Yellow, Green and Blue (RYGB) which are produced by laser diodes (LDs) and are mixed to produce white light). While the second scenario considered is the best scenario where users are distributed over the room and each access point just serves one user. The optical channel bandwidth was determined from the impulse response using the parameters in Table 1. Figs. 2 and 3 show the CDF of the channel bandwidth of 32 locations inside the room using ADR and ImR respectively.

The controller which is located on the room ceiling (See Fig. 1), is assumed to have users location similar to the work in [38]. The performance of the VLC system can be enhanced by optimising the access points allocation and wavelength resources allocation so as to minimise interference and maximise the sum SINR.

A MILP model similar to [38], [39] was developed to maximise the sum of SINRs of all users based on wavelengths assignment optimisation. Note that [38] considered a wide field of view receiver, [39] considered an ADR. This work shows the performance improvement that can be obtained by using an imaging receiver, where the results are compared to those obtained using an ADR. The powers received at 32 different locations inside the room from each AP using different wavelengths have been precalculated using the ray tracing tool and hence the impulse responses at these locations. These results are provided to the MILP model. The optimised assignment of AP and wavelength to each user is determined by the MILP model based on maximising the sum of SINRs. Fig. 1 shows an illustrative example of how the system works based on a scenario that has three users. The assignment of an access point and a wavelength that carries modulated data to a user is illustrated using solid lines. While the interference between users using the same wavelength is indicated by dashed lines and finally dotted lines show unmodulated wavelength at an access point, which are used purely for illumination and hence act as background noise. User 1 in the example in Fig. 1 suffers from background noise only, as the blue wavelength is not assigned to other users. While users 2 and 3 suffer from interference as both of them are assigned to the same wavelength (red wavelength) in addition to the background noise.

The SINR of a user ($u$) who is assigned to an access point ($a$) with wavelength ($\lambda$) using pixel ($p$) in the imaging receiver, can be written as follows:

$$SINR_{u,p}^{a,\lambda} = \frac{Signal}{Interference + Noise} = \frac{RP_{u,p}^{a,\lambda} S_{u,p}^{a,\lambda}}{\sum_{\substack{b \in \mathcal{A} \\ b \neq a}} \sum_{\substack{m \in \mathcal{U} \\ m \neq u}} \sum_{i \in P} RP_{u,p}^{b,\lambda} S_{m,p}^{b,\lambda} + \sum_{\substack{b \in \mathcal{A} \\ b \neq a}} \sigma_{u,p}^{b,\lambda} \left[1 - \sum_{\substack{m \in \mathcal{U} \\ m \neq u}} \sum_{i \in \mathcal{B}} S_{m,i}^{b,\lambda}\right] + \sigma_{Rx}}$$

where $R$ is the receiver responsivity in A/W, and $RP_{u,p}^{a,\lambda}$ is the received electrical current (signal) by user $u$ assigned to access point $a$ and wavelength $\lambda$ and using receiver pixel $p$, $S_{u,p}^{a,\lambda}$ is a binary assignment function that equal to 1 if user $u$ is assigned to access point $a$ and wavelength $\lambda$ using pixel $p$ at the receiver, $RP_{u,p}^{b,\lambda}$ is the received electrical current from another access point, $b$, (interference), $\sigma_{u,p}^{b,\lambda}$ is the background noise and $\sigma_{Rx}$ is the receiver noise.

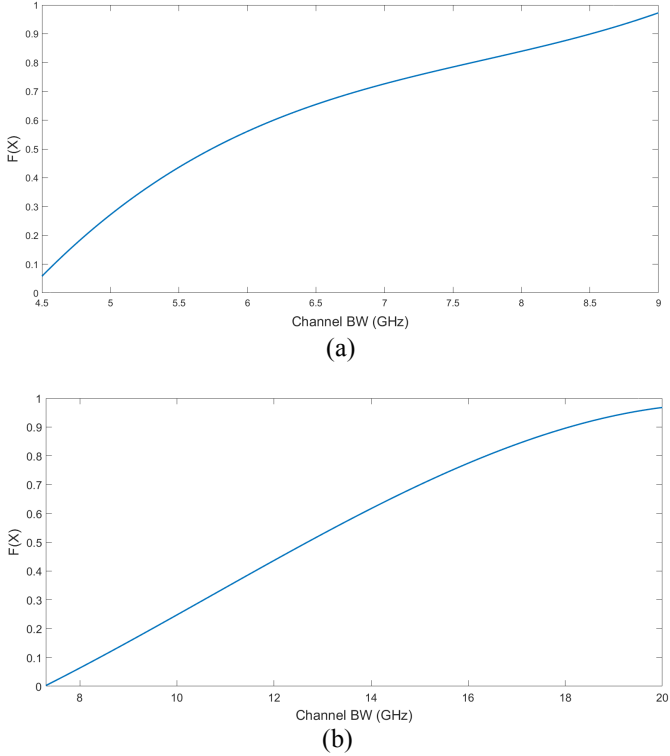

**Fig. 2.** CDF of the optical channel bandwidth at different locations in the room for (a) ADR, (b) ImR.

**Table 1.** Room Configurations

| Parameters | Configurations | |
|---|---|---|
| Room | | |
| Length × Width × Height | 8 m × 4 m × 3 m | |
| Walls and Ceiling reflection coefficient | 0.8 [37] | |
| Floor reflection coefficient | 0.3 [37] | |
| Number of Reflections | 1 | 2 |
| Area of reflection element | 5 cm × 5 cm | 20 cm × 20 cm |
| Order of Lambertian pattern, walls, floor and ceiling | 1 [37] | |
| Semi-angle of reflection element at half power | 60° | |

| | |
|---|---|
| Transmitters | |
| Number of transmitters' units | 8 |
| Transmitters locations (x, y, z) | (1 m, 1 m, 3 m), (1 m, 3 m, 3 m), (1 m, 5 m, 3 m), (1 m, 7 m, 3 m), (3 m, 1 m, 3 m), (3 m, 3 m, 3 m), (3 m, 5 m, 3 m) and (3 m, 7 m, 3 m) |
| Number of RYGB LDs per unit | 12 |
| Transmitted optical power of Red LD | 0.8 W |
| Transmitted optical power of Yellow LD | 0.5 W |
| Transmitted optical power of Green LD | 0.3 W |
| Transmitted optical power of Blue LD | 0.3 W |
| Total transmitted power of RYGB LD | 1.9 W |
| Semi-angle at half power | 60° |
| Receiver | |
| Responsivity Red | 0.4 A/W |
| Responsivity Yellow | 0.35 A/W |
| Responsivity Green | 0.3 A/W |
| Responsivity Blue | 0.2 A/W |
| ADR | |
| Number of Photodetectors | 4 |
| Area of the photodetector | 20 mm² |
| Photodetector | 1    2    3    4 |
| Azimuth angels | 45°  135°  225°  315° |
| Elevation angels | 70°  70°  70°  70° |
| Field of view (FOV) of each detector | 25° |
| Receiver noise current spectral density | 4.47 pA/ √ Hz [35] |
| Receiver bandwidth | 5 GHz |
| ImR | |
| Number of Pixels | 9 |
| Area of the photodetector | 16 mm² |
| Field of view (FOV) of lens | 50° |
| Receiver noise current spectral density | 10 pA/√Hz [40] |
| Receiver bandwidth | 10 GHz |

**Table 2.** Scenario 1 with the optimized resource allocation

| User | Location (x,y,z) | ADR | | | ImR | | |
|---|---|---|---|---|---|---|---|
| | | AP | Branch | wavelength | AP | Pixel | wavelength |
| 1 | (0.5,6.5,1) | 3 | 4 | Red | 3 | 4 | Red |
| 2 | (0.5,7.5,1) | 4 | 4 | Yellow | 4 | 5 | Red |
| 3 | (1.5,6.5,1) | 8 | 1 | Red | 4 | 5 | Yellow |
| 4 | (1.5,7.5,1) | 4 | 3 | Red | 8 | 8 | Red |
| 5 | (2.5,0.5,1) | 5 | 1 | Red | 1 | 2 | Red |
| 6 | (2.5,1.5,1) | 1 | 3 | Red | 5 | 5 | Yellow |
| 7 | (3.5,0.5,1) | 5 | 2 | Yellow | 5 | 5 | Red |
| 8 | (3.5,1.5,1) | 6 | 2 | Red | 6 | 6 | Red |

**Table 3.** Scenario 2 with the optimized resource allocation.

| User | Location (x,y,z) | ADR | | | ImR | | |
|---|---|---|---|---|---|---|---|
| | | AP | Branch | wavelength | AP | Pixel | wavelength |
| 1 | (0.5,5.5,1) | 3 | 4 | Red | 3 | 5 | Red |
| 2 | (1.5,1.5,1) | 1 | 3 | Red | 1 | 5 | Red |
| 3 | (1.5,3.5,1) | 2 | 3 | Red | 2 | 5 | Red |
| 4 | (1.5,7.5,1) | 4 | 3 | Red | 4 | 5 | Red |
| 5 | (2.5,2.5,1) | 6 | 1 | Red | 6 | 5 | Red |
| 6 | (2.5,6.5,1) | 8 | 1 | Red | 8 | 5 | Red |
| 7 | (3.5,1.5,1) | 5 | 3 | Red | 5 | 5 | Red |
| 8 | (3.5,4.5,1) | 7 | 2 | Red | 7 | 5 | Red |

The optical channel bandwidth, the SINR and achievable data rates have been evaluated for each user in each scenario in this work after the assignment of APs and wavelengths is optimised using the MILP model. The results are shown in Figs. (4-7). Three factors can limit the achievable data rate: the light source modulation bandwidth, the channel bandwidth and/or the receiver bandwidth. In this work, laser diodes (LDs) were used as light source (with diffusers to expand the beam and provide eye safe operation [35]) which can support high data rates in the Gbps and beyond [35]. Also, ADR and ImR were used to improve the channel bandwidth by limiting the interference and background noise. The limited field of view of each face (ADR) / pixel (ImR) decrease the delay spread, and hence increases the channel bandwidth seen by the system. The achievable data rates in this work are limited however by the receiver bandwidth when using the ADR where the supported data rate is 7.1 Gbps per user, while, the ImR can support 14.2 Gbps per user.

The channel bandwidth is evaluated using the user and AP locations and the propagation environment. The results show a minimum channel bandwidth of 4.5 GHz and a maximum of 9 GHz when using ADR (See Fig. 2a), while, the minimum channel bandwidth is around 7.5 GHz and the maximum is 20 GHz when using ImR (See Fig. 2b). The ImR provides a better channel bandwidth compared to the ADR for both scenarios as shown in Figs. 4 and 6. The ImR has a narrower field of view per pixel which limits the range of rays accepted and hence reduces the delay spread, thus increasing the channel bandwidth. The use of multiple receiver pixels in the ImR can help increase the amount of collected optical power. The SINR was calculated for both scenarios based on the results of the optimised resource allocations. When using ADR, the SINR was examined at a data rate of 7.1 Gbps; while when utilising the ImR, the SINR was evaluated at a data rate 14.2 Gbps. The ImR provides a higher SINR than the ADR as shown in Figs. 5 and 7.

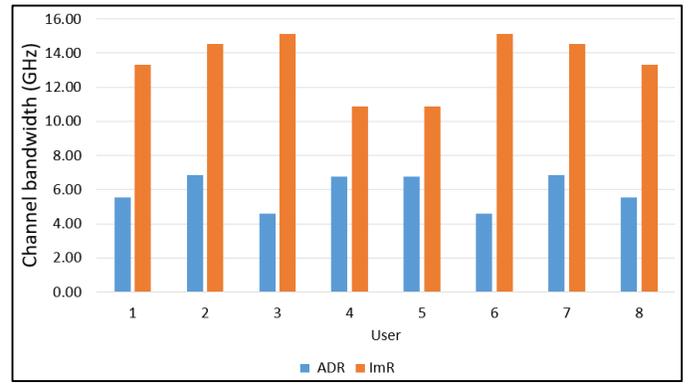

**Fig. 4.** Optical channel bandwidth in Scenario 1.

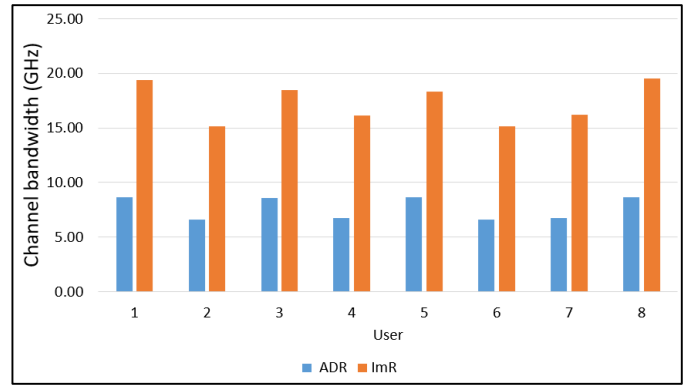

**Fig. 5.** SINR for different users in Scenario 1.

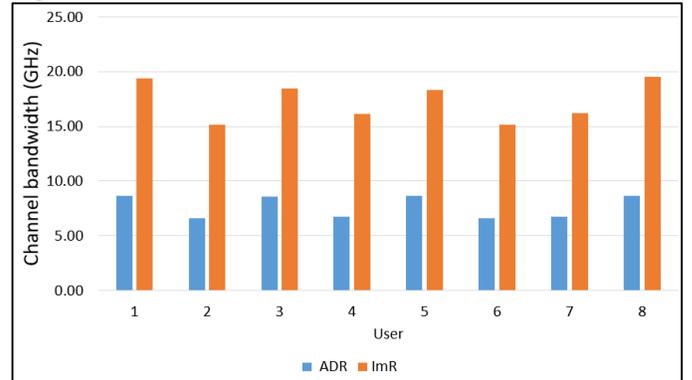

**Fig. 6.** Optical channel bandwidth in Scenario 2.

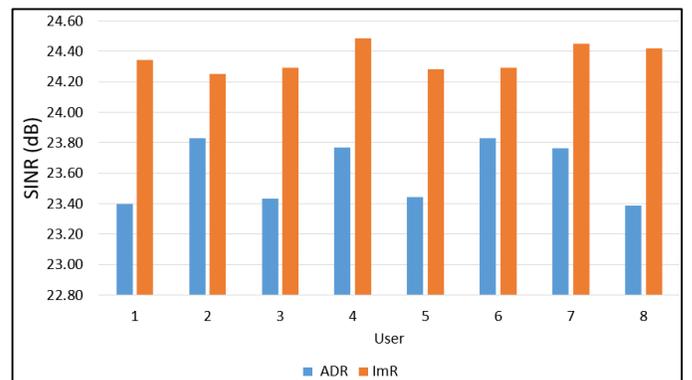

**Fig. 7.** SINR for different users in Scenario 2.

## IV. Conclusions

A wavelength division multiple access (WDMA) scheme was evaluated in this work to support multiple high data rate users in an indoor environment. A mixed-integer linear programming (MILP) model was proposed to optimise the resource allocation in optical wireless communication systems. In addition, two types of receivers were used and compared in this work, namely an angle diversity receiver (ADR) and an imaging receiver (ImR). The ImR can provide a higher data rate of up to 14.2 Gbps to each user with high signal to interference and noise ratio (SINR), while the ADR can support a data rate of 7.1 Gbps. The ImR receiver offers a better result when compared to the ADR in terms of channel bandwidth, SINR and data rate. The use of infrared which can support up to 100 wavelengths and beyond (our VLC system had 4 wavelengths), can increase the aggregate capacity provided by the access points. Spatial multiplexing can also be used, with the resulting WDM infrared system having the potential to realise Tb/s data rates.

## Acknowledgments

The authors would like to acknowledge funding from the Engineering and Physical Sciences Research Council (EPSRC) INTERNET (EP/H040536/1), STAR (EP/K016873/1) and TOWS (EP/S016570/1) projects. The authors extend their appreciation to the deanship of Scientific Research under the International Scientific Partnership Program ISPP at King Saud University, Kingdom of Saudi Arabia for funding this research work. OZA would like to thank Umm Al-Qura University in the Kingdom of Saudi Arabia for funding his PhD scholarship. AAA would like to thank the Imam Abdulrahman Bin Faisal University in the Kingdom of Saudi Arabia for funding her PhD scholarship, SOMS would like to thank the University of Leeds and the Higher Education Ministry in Sudan for funding her PhD scholarship. SHM would like to thank EPSRC for providing her Doctoral Training Award scholarship. All data are provided in full in the results section of this paper.

## References

[1] A. Alexiou and M. Haardt, "Smart antenna technologies for future wireless systems: Trends and challenges," *IEEE Commun. Mag.*, vol. 42, no. 9, pp. 90–97, 2004.

[2] A. J. Paulraj, D. A. Gore, R. U. Nabar, and H. Bölcskei, "An overview of MIMO communications - A key to gigabit wireless," in *Proceedings of the IEEE*, 2004, vol. 92, no. 2, pp. 198–217.

[3] Cisco Mobile, *Cisco Visual Networking Index: Global Mobile Data Traffic Forecast Update, 2016-2021 White Paper*. 2017.

[4] Z. Ghassemlooy, W. Popoola, and S. Rajbhandari, *Optical wireless communications: system and channel modelling with Matlab®*. 2012.

[5] F. E. Alsaadi, M. A. Alhartomi, and J. M. H. Elmirghani, "Fast and efficient adaptation algorithms for multi-gigabit wireless infrared systems," *J. Light. Technol.*, vol. 31, no. 23, pp. 3735–3751, 2013.

[6] A. T. Hussein and J. M. H. Elmirghani, "10 Gbps Mobile Visible Light Communication System Employing Angle Diversity, Imaging Receivers, and Relay Nodes," *J. Opt. Commun. Netw.*, vol. 7, no. 8, pp. 718–735, 2015.

[7] S. H. Younus and J. M. H. Elmirghani, "WDM for high-speed indoor visible light communication system," in *International Conference on Transparent Optical Networks*, 2017, pp. 1–6.

[8] A. T. Hussein, M. T. Alresheedi, and J. M. H. Elmirghani, "20 Gb/s Mobile Indoor Visible Light Communication System Employing Beam Steering and Computer Generated Holograms," *J. Light. Technol.*, vol. 33, no. 24, pp. 5242–5260, 2015.

[9] A. T. Hussein, M. T. Alresheedi, and J. M. H. Elmirghani, "25 Gbps mobile visible light communication system employing fast adaptation techniques," in *2016 18th International Conference on Transparent Optical Networks (ICTON)*, 2016.

[10] O. Z. Alsulami, M. T. Alresheedi, and J. M. H. Elmirghani, "Transmitter diversity with beam steering," in *2019 21st International Conference on Transparent Optical Networks (ICTON)*, 2019, pp. 1–5.

[11] O. Z. Alsulami, M. O. I. Musa, M. T. Alresheedi, and J. M. H. Elmirghani, "Visible light optical data centre links," in *2019 21st International Conference on Transparent Optical Networks (ICTON)*, 2019, pp. 1–5.

[12] O. Z. Alsulami, M. T. Alresheedi, and J. M. H. Elmirghani, "Optical Wireless Cabin Communication System," in *2019 IEEE Conference on Standards for Communications and Networking (CSCN)*, 2019, pp. 1–4.

[13] O. Z. Alsulami, M. O. I. Musa, M. T. Alresheedi, and J. M. H. Elmirghani, "Co-existence of Micro, Pico and Atto Cells in Optical Wireless Communication," in *2019 IEEE Conference on Standards for Communications and Networking (CSCN)*, 2019, pp. 1–5.

[14] K. L. Sterckx, J. M. H. Elmirghani, and R. A. Cryan, "Sensitivity assessment of a three-segment pyrimadal fly-eye detector in a semi-disperse optical wireless communication link," *IEE Proc. Optoelectron.*, vol. 147, no. 4, pp. 286–294, 2000.

[15] A. G. Al-Ghamdi and J. M. H. Elmirghani, "Performance evaluation of a triangular pyramidal fly-eye diversity detector for optical wireless communications," *IEEE Commun. Mag.*, vol. 41, no. 3, pp. 80–86, 2003.

[16] M. T. Alresheedi and J. M. H. Elmirghani, "Hologram Selection in Realistic Indoor Optical Wireless Systems With Angle Diversity Receivers," *IEEE/OSA J. Opt. Commun. Netw.*, vol. 7, no. 8, pp. 797–813, 2015.

[17] A. T. Hussein, M. T. Alresheedi, and J. M. H. Elmirghani, "Fast and Efficient Adaptation Techniques for Visible Light Communication Systems," *J. Opt. Commun. Netw.*, vol. 8, no. 6, pp. 382–397, 2016.

[18] S. H. Younus, A. A. Al-Hameed, A. T. Hussein, M. T. Alresheedi, and J. M. H. Elmirghani, "Parallel Data Transmission in Indoor Visible Light Communication Systems," *IEEE Access*, vol. 7, pp. 1126–1138, 2019.

[19] A. G. Al-Ghamdi and M. H. Elmirghani, "Optimization of a triangular PFDR antenna in a fully diffuse OW system influenced by background noise and multipath propagation," *IEEE Trans. Commun.*, vol. 51, no. 12, pp. 2103–2114, 2003.

[20] A. G. Al-Ghamdi and J. M. H. Elmirghani, "Characterization of mobile spot diffusing optical wireless systems with receiver diversity," in *ICC'04 IEEE International Conference on Communications*, 2004.

[21] F. E. Alsaadi and J. M. H. Elmirghani, "Performance evaluation of 2.5 Gbit/s and 5 Gbit/s optical wireless systems employing a two dimensional adaptive beam clustering method and imaging diversity detection," *IEEE J. Sel. Areas Commun.*, vol. 27, no. 8, pp. 1507–1519, 2009.

[22] M. T. Alresheedi and J. M. H. Elmirghani, "10 Gb/s indoor optical wireless systems employing beam delay, power, and angle adaptation methods with imaging detection," *IEEE/OSA J. Light. Technol.*, vol. 30, no. 12, pp. 1843–1856, 2012.

[23] K. L. Sterckx, J. M. H. Elmirghani, and R. A. Cryan, "Pyramidal fly-eye detection antenna for optical wireless systems," *Opt. Wirel. Commun. (Ref. No. 1999/128), IEE Colloq.*, p. 5/1-5/6, 1999.

[24] F. E. Alsaadi, M. Nikkar, and J. M. H. Elmirghani, "Adaptive mobile optical wireless systems employing a beam clustering method, diversity detection, and relay nodes," *IEEE Trans. Commun.*, vol. 58, no. 3, pp. 869–879, 2010.

[25] F. E. Alsaadi and J. M. H. Elmirghani, "Adaptive mobile line strip multibeam MC-CDMA optical wireless system employing imaging detection in a real indoor environment," *IEEE J. Sel. Areas Commun.*, vol. 27, no. 9, pp. 1663–1675, 2009.

[26] M. T. Alresheedi and J. M. H. Elmirghani, "Performance evaluation of 5 Gbit/s and 10 Gbit/s mobile optical wireless systems employing beam angle and power adaptation with diversity receivers," *IEEE J. Sel. Areas Commun.*, vol. 29, no. 6, pp. 1328–1340, 2011.

[27] F. E. Alsaadi and J. M. H. Elmirghani, "Mobile Multi-gigabit Indoor Optical Wireless Systems Employing Multibeam Power Adaptation and Imaging Diversity Receivers," *IEEE/OSA J. Opt. Commun. Netw.*, vol. 3, no. 1, pp. 27–39, 2011.

[28] A. G. Al-Ghamdi and J. M. H. Elmirghani, "Line Strip Spot-Diffusing Transmitter Configuration for Optical Wireless Systems Influenced by Background Noise and Multipath Dispersion," *IEEE Trans. Commun.*, vol. 52, no. 1, pp. 37–45, 2004.

[29] F. E. Alsaadi and J. M. H. Elmirghani, "High-speed spot diffusing mobile optical wireless system employing beam angle and power adaptation and imaging receivers," *J. Light. Technol.*, vol. 28, no. 16, pp. 2191–2206, 2010.

[30] G. Cossu, a M. Khalid, P. Choudhury, R. Corsini, and E. Ciaramella, "3.4 Gbit/s visible optical wireless transmission based on RGB LED.," *Opt. Express*, vol. 20, no. 26, pp. B501-6, 2012.

[31] Y. Wang, Y. Wang, N. Chi, J. Yu, and H. Shang, "Demonstration of 575-Mb/s downlink and 225-Mb/s uplink bi-directional SCM-WDM visible light communication using RGB LED and phosphor-based LED," *Opt. Express*, vol. 21, no. 1, p. 1203, 2013.

[32] A. Neumann, J. J. Wierer, W. Davis, Y. Ohno, S. R. J. Brueck, and J. Y. Tsao, "Four-color laser white illuminant demonstrating high color-rendering quality," *Opt. Express*, vol. 19, no. S4, p. A982, 2011.

[33] F.-M. Wu, C.-T. Lin, C.-C. Wei, C.-W. Chen, Z.-Y. Chen, and K. Huang, "3.22-Gb/s WDM Visible Light Communication of a Single RGB LED Employing Carrier-Less Amplitude and Phase Modulation," in *Optical Fiber Communication Conference/National Fiber Optic Engineers Conference 2013*, 2013, p. OTh1G.4.

[34] T. A. Khan, M. Tahir, and A. Usman, "Visible light communication using wavelength division multiplexing for smart spaces," *Consumer Communications and Networking Conference (CCNC), 2012 IEEE*. pp. 230–234, 2012.

[35] A. T. Hussein and J. M. H. Elmirghani, "Mobile Multi-Gigabit Visible Light Communication System in Realistic Indoor Environment," *J. Light. Technol.*, vol. 33, no. 15, pp. 3293–3307, 2015.

[36] J. R. Barry, J. M. Kahn, W. J. Krause, E. A. Lee, and D. G. Messerschmitt, "Simulation of Multipath Impulse Response for Indoor Wireless Optical Channels," *IEEE J. Sel. Areas Commun.*, vol. 11, no. 3, pp. 367–379, 1993.

[37] F. R. Gfeller and U. Bapst, "Wireless In-House Data Communication via Diffuse Infrared Radiation," *Proc. IEEE*, vol. 67, no. 11, pp. 1474–1486, 1979.

[38] S. O. M. Saeed, S. Hamid Mohamed, O. Z. Alsulami, M. T. Alresheedi, and J. M. H. Elmirghani, "Optimized resource allocation in multi-user WDM VLC systems," in *2019 21st International Conference on Transparent Optical Networks (ICTON)*, 2019, pp. 1–5.

[39] O. Z. Alsulami, A. A. Alahmadi, S. O. M. Saeed, S. Hamid Mohamed, T. E. H. El-Gorashi, M. T. Alresheedi, and J. M. H. Elmirghani, "Optimum resource allocation in optical wireless systems with energy efficient fog and cloud architectures," *Philos. Trans. R. Soc. A Math. Phys. Eng. Sci.*, to be published.

[40] E. Kimber, B. Patel, I. Hardcastle, and A. Hadjifotiou, "High performance 10 Gbit/s pin-FET optical receiver," *Electron. Lett.*, vol. 28, no. 2, pp. 120–122, 1992.